\documentclass{aastex61}
\usepackage{listings}

\begin{document}

\title{High Performance Negative Database for Massive Data Management System of The Mingantu Spectral Radioheliograph}

\correspondingauthor{Feng Wang}
\email{wangfeng@acm.org}

\author[0000-0002-0786-7307]{Congming Shi}
\affil{Faculty Of Management And Economics, Key Lab of Computer Technology Application of Yunnan Province,\\
Kunming University of Science and Technology \\
747 Jingming Nan Rd, Chenggong Campus, Kunming \\
Yunnan, China,650500}

\author{Feng Wang}
\affil{Faculty Of Management And Economics, Key Lab of Computer Technology Application of Yunnan Province,\\
Kunming University of Science and Technology \\
747 Jingming Nan Rd, Chenggong Campus, Kunming \\
Yunnan, China,650500}
\affiliation{Yunnan Observatories, Chinese Academy of Sciences, Kunming, Yunnan, 650011}


\author{Hui Deng}
\affiliation{Key Lab of Computer Technology Application of Yunnan Province, \\
Kunming University of Science and Technology \\
747 Jingming Nan Rd, Chenggong Campus, Kunming \\
Yunnan, China,650500}

\author{Yingbo Liu}
\affiliation{Yunnan academy of scientific $\&$ technical information, Kunming
Yunnan, China,650051}

\author{Cuiyin Liu}
\affiliation{Key Lab of Computer Technology Application of Yunnan Province, \\
Kunming University of Science and Technology \\
747 Jingming Nan Rd, Chenggong Campus, Kunming \\
Yunnan, China,650500}

\author{Shoulin Wei}
\affiliation{Key Lab of Computer Technology Application of Yunnan Province, \\
Kunming University of Science and Technology \\
747 Jingming Nan Rd, Chenggong Campus, Kunming \\
Yunnan, China,650500}
\nocollaboration



\begin{abstract}
As a dedicated synthetic aperture radio interferometer, the MingantU SpEctral Radioheliograph (MUSER), initially known as the Chinese Spectral RadioHeliograph (CSRH), has entered the stage of routine observation. More than 23 million data records per day need to be effectively managed to provide high  performance data query and retrieval for scientific data reduction. In light of these  massive amounts of data generated by the MUSER, in this paper, a novel data management technique called the negative database (ND) is proposed and used to implement a data management system for the MUSER. Based on the key-value database, the ND technique makes complete utilization of the complement set of observational data to derive the requisite information. 
Experimental results showed that the proposed ND can significantly reduce storage volume in comparison with a relational database management system (RDBMS). Even when considering the time needed to derive records that were absent, its overall performance, including querying and deriving the data of the ND, is comparable with that of an RDBMS.
The ND technique effectively solves the problem of massive data storage for the MUSER, and is a valuable reference for the massive data management required in next-generation telescopes.

\end{abstract}

\keywords{massive data management, MUSER--astronomical instrument}



\section{Introduction}
\label{sec:introduction}
The MingantU SpEctral Radioheliograph (MUSER) is a synthetic aperture radio interferometer capable of observing radio bursts and producing high-quality radio images at frequencies ranging from 0.4 GHz to 15 GHz with high temporal, spatial, and spectral resolution. The MUSER is divided into a low-frequency sub-array (MUSER-I) and a high-frequency sub-array (MUSER-II) \citep{yan2004chinese,yan2009chinese}. A total of 100 radio antennas of the MUSER are spirally distributed in Ming’antu, Inner Mongolia, in China. The specifications of the MUSER are listed in Table~\ref{tblSpecification}.

\begin{table*}[!htb]
\small
\centering
\caption{The specifications of MUSER\label{tblSpecification}}
\begin{tabular}{lll}
\hline  \hline
 & MUSER-I & MUSER-II \\
\hline
Frequency range & 0.4-2 GHz &  2 - 15 GHz \\
Antennas & 40 &  60 \\
Baselines & 780 &  1770 \\
IF Bandwidth & 25 MHz & 25MHz \\
Band & 4 (0.4, 0.8, 1.2, 1.6 GHz) & 33 (2,2.4, 2.8, ..., 14.8 GHz) \\
Frequency resolution & 64 channels & 528 channels\\
Temporal resolution &  25 ms (3ms/frame) & 206.25 ms  \\
Polarizations & Dual circular L, R & Dual circular L, R \\
Lmax & app. 3 km & app. 3 km\\
\hline
\end{tabular}
\end{table*}

The MUSER will generate massive amounts of observational data. The digital receiver of the MUSER I/II sends a frame to a data acquisition server through a 1.25 Gb optical fiber every 3.125 ms. A frame contains visibility data and the autocorrelation of 16 channels with one polarization. Therefore, the digital receivers will output multiple consecutive frames for data with all channels and all polarizations. The MUSER-I takes 25 ms ($3.125 ms\times8$) to generate eight consecutive frames treated as a full frame (64 channels and two polarizations). The size of a frame in MUSER-I is 100,000 bytes.
The MUSER-II takes 206.25 ms ($3.125 ms\times66$) to generate 66 consecutive frames treated as a full frame (528 channels and two polarizations). The size of a frame in MUSER-II is 204,800 bytes. In an observational day consisting of 10 hours, the MUSER-I will produce 11.52 million data frames and the MUSER-II 11.52 million data frames.

\begin{figure*}[htbp]
\centering
\includegraphics[width=0.6\textwidth]{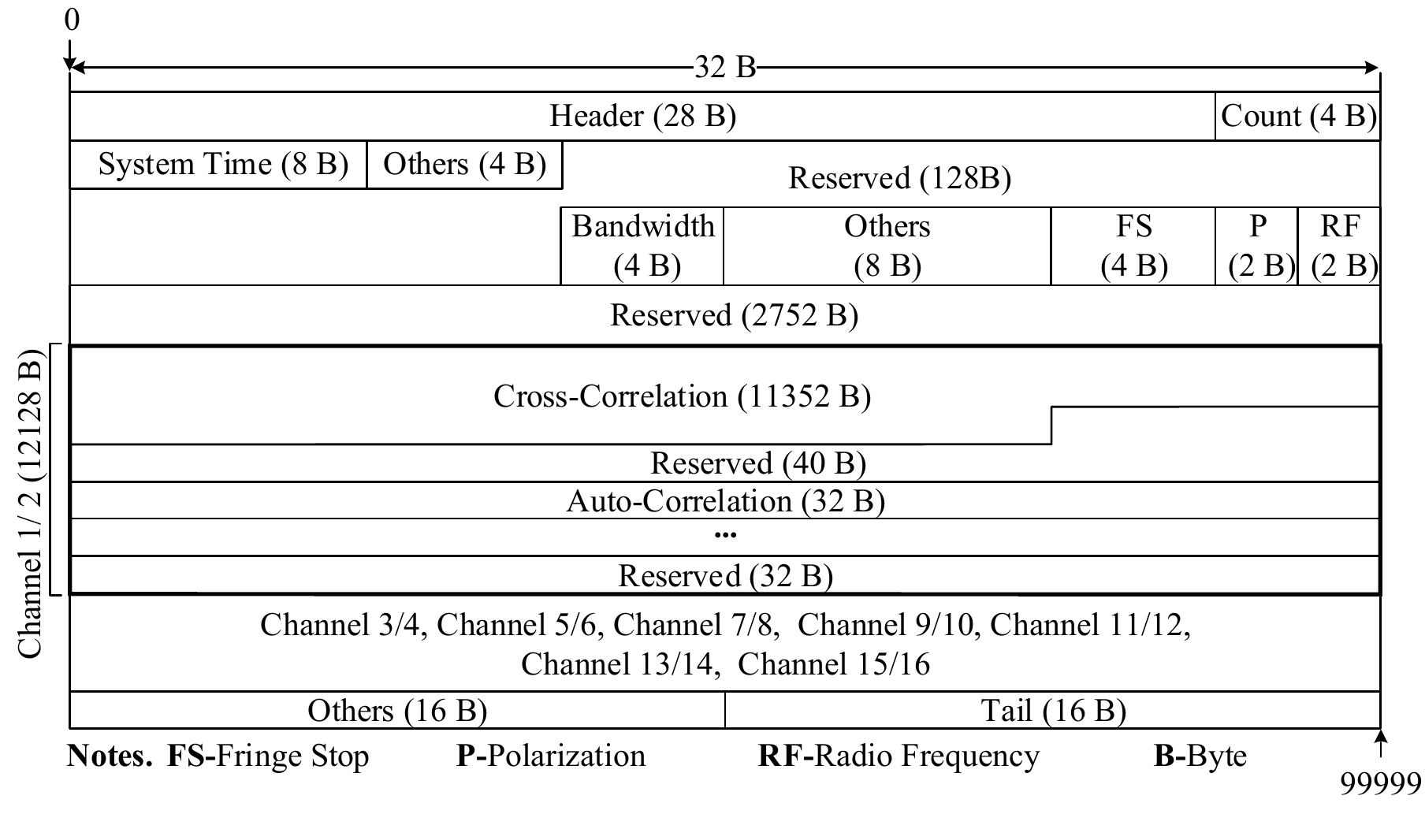}
\caption{MUSER-I data format. Cite from~\citep{wang2015distributed}}\label{Fig-MUSER-I-Frame}
\end{figure*}

\begin{figure*}[htbp]
\centering
\includegraphics[width=0.6\textwidth]{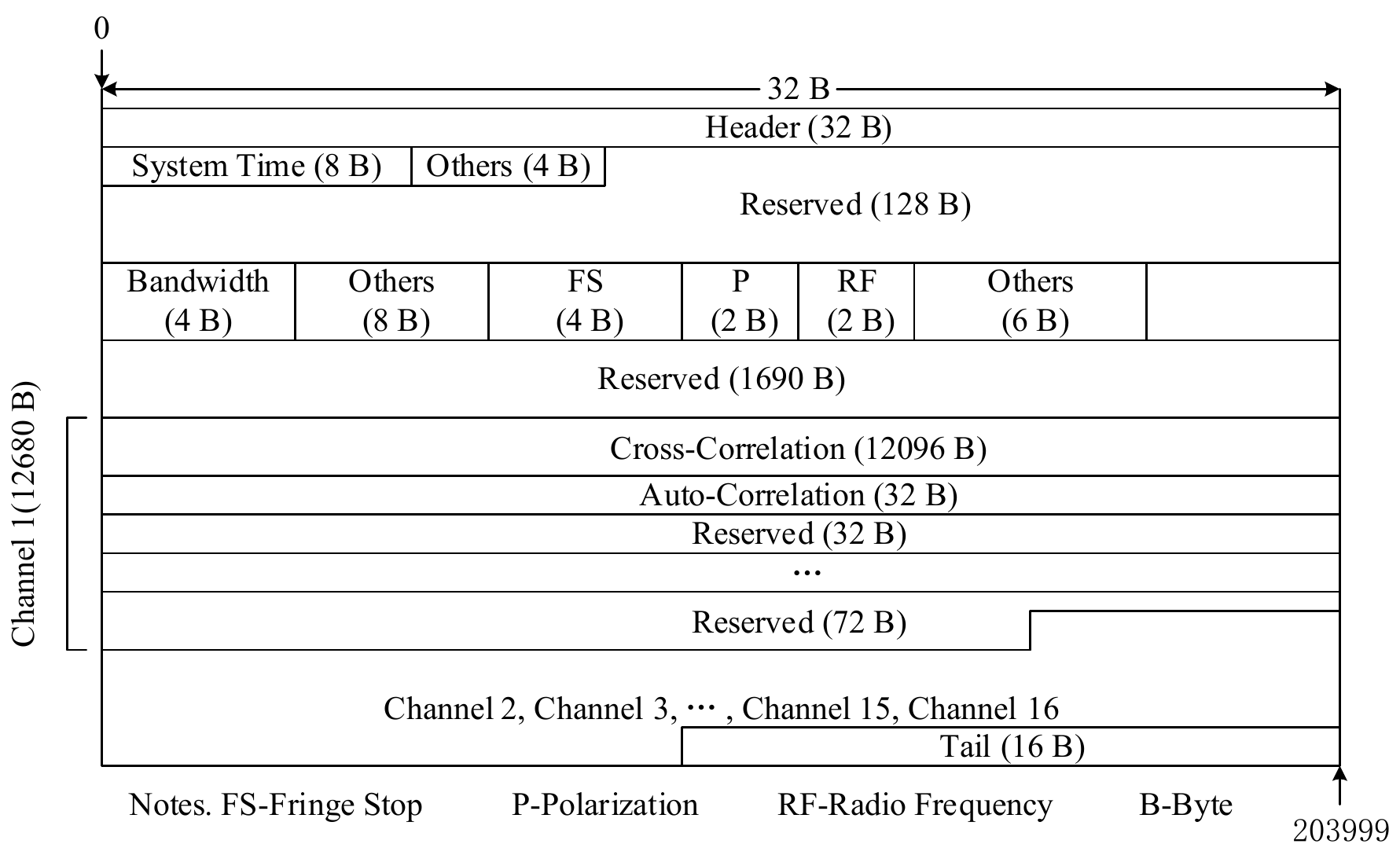}
\caption{MUSER-II data format.}\label{Fig-MUSER-II-Frame}
\end{figure*}

Useful observational information is encapsulated in a frame (the frame format of the MUSER-I is shown in  Figure~\ref{Fig-MUSER-I-Frame} and that of the MUSER-II in Figure~\ref{Fig-MUSER-II-Frame}). In addition to visibility data and autocorrelation, many important observational parameters (metadata), such as observational date, time, band, and polarization, can be read out. 


To manage such massive amounts of data produced by the MUSER, it is necessary to design a data management system (DMS)~\citep{wang2015distributed} to store the metadata of each frame so that the data processing pipeline can quickly retrieve and locate the specified frame(s). The DMS is very important because almost all the main functions of the MUSER data processing pipeline, such as UVFITS file generation, imaging, and data integration, are heavily dependent on the DMS.

In this paper, we focus on designing and implementing a high performance data management system for the MUSER. The remainder of this paper is organized as follows. In Section 2, we briefly overview related work. In Section 3, we introduce the MUSER data management system. Section 4 describes a comprehensive set of experiments performed to verify various performance-related aspects of our proposal. Section 5 provides a discussion of the ND technique. Finally, Section 6 contains the conclusions of this study.

\section{Related Work} 

Almost every modern telescope system has at least one storage system to store telescope operations and observational data in general. A database system, often called the data management system, the catalog system, or the index system, is also setup and maintained to store metadata to provide data retrieval services. 

Database technology has been widely applied to astronomy. All past work in the area that has discussed data production, survey catalogs, and ephemeris contains detailed descriptions of methods to construct databases. In recent years, more advanced database technologies have been introduced to astronomical research. Therefore, we review related work from two aspects to clearly understand the development of database technology for astronomical data management. 

1. Relational database technology

A number of classical, mature commercial or open-source database management systems (RDBMS), such as Oracle, MySQL, Microsoft SQL Server, PostgreSQL, DB2, and so on, have been widely used for astronomical data management. 

One of the best known cases of astronomical data management is the data management system for the Sloan Digital Sky Survey (SDSS). Microsoft's SQL Server was applied with the collaboration of the SDSS to maintain the SDSS data archive and the SkyServer website, which provided a range of interfaces to an underlying Microsoft SQL Server, made SDSS data available to astronomers, teachers, and citizen scientists \citep{szalay2002sdss,gray2002data,szalay2008sloan,raddick2014ten}.

Similar to SDSS, the data cycle management system of the Large Sky Area Multi-Object Fibre Spectroscopic Telescope (LAMOST) is based on RDBMS.  The PostgreSQL database and pgSphere were used to manage the LAMOST data archive and data release \citep{cui2012large,he2016lamost}.

SciDB is a commercial, open-source analytical array database oriented toward the data management needs of scientists, and has been applied to the Large Synoptic Survey Telescope (LSST) \citep{stonebraker2011architecture}. The millions of light curves for the LOw-Frequency ARray (LOFAR) were managed by the MonetDB/SQL \& SciQL on their SciLens platform \citep{scheers2012towards}. The LOFAR long-term archive (LTA) is a distributed information system that was created to store and process large volumes of data generated by the LOFAR radio telescope. Most standard LOFAR data products are stored and managed using HDF5 \citep{alexov2012status,van2013lofar}.

In addition to deployment in survey projects, RDBMSs have been used in data processing systems. For example, Astro-WISE is an integrated data-centric system where processing, storage, and administration is integrated into a single environment, providing a living system to both data producers and the customers, the data model for which was implemented both in the relational database Oracle 11g RAC and the hierarchy of Python classes, and the database for which was partitioned into instruments and projects \citep{valentijn2007astro,begeman2013astro}. 

2. NoSQL and other technologies

Since its advent, Not Only Structured Query Language (NoSQL) technology, such as Fastbit, has been intensively studied and applied to massive astronomical data processing, especially to massive data distributed storage and data index/query. 

FastBit is a software tool for searching read-only datasets and locating the required information from massive amounts of data produced by scientific instruments and computer simulations. It utilizes a compressed bitmap index, encoding, and binning \citep{wu2005fastbit,wu2009fastbit}. A new data archiving system for the NVST based on the Fastbit NoSQL database has been implemented, where the average number of records generated by NVST in one year is approximately 10 to 12 million \citep{liu2014nvst}.
A Distributed Data Storage System (DDSS) for real-time observational data storage based on the Lustre distributed file system was implemented on the NVST \citep{liu2015low}.

Redis is among the most widely used in-memory data structure databases that can support several types of data structures, such as hashes, sets, and sorted sets with range queries. It is used to test the performance of AQUAdexIM, which is an innovative spatial indexing and querying method for massive time-series images \citep{hong2016aquadexim}.

In summary, the above survey shows that while a number of studies in the literature have discussed the implementation of the DMS using database systems, only a few have considered systems to manage data at a scale comparable to that generated by the MUSER. It is thus useful to determine methods to create a data management system that delivers high performance using low storage volume, as this is needed for modern telescope systems. 

\section{MUSER Data Management System}

\subsection{Requirements Analysis}
\begin{figure*}[htbp]
   \centering
   \includegraphics[width=0.8\linewidth]{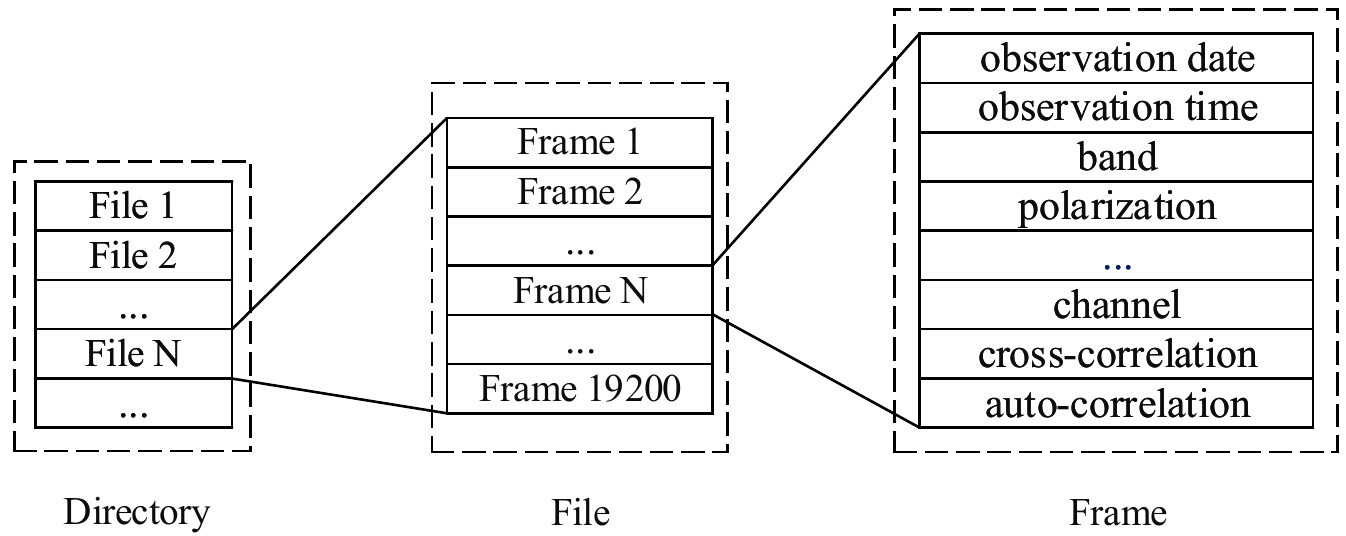}
   \caption{ Data storage organization. }\label{FigureDatOrg}
\end{figure*}

Figure~\ref{FigureDatOrg} shows a diagram of data storage organization for the current MUSER data system. All observational data are stored in the storage system in the form of files. A total of 19,200 consecutive frames captured in one minute are stored in a file. The file name is created as ``YYYYMMDD-hhmm'' according to the time of the first frame in the file. The files of the MUSER-I and the MUSER-II are stored in different directories. 

Two situations arising during observation require special consideration. 1) A few frames are sometimes lost. This is mainly due to the performance of currently available storage systems. The lost frames are completely random.  2) The starting time of observation every day is determined by the observer. The time of the first observation in a day is when the observer starts observation. Therefore, it is challenging to start observation at 0 seconds of a minute. This means that the observational data recorded in a minute may be stored in two files. For example, if the observer starts the observation at 08h:30m:30s, the observational data for 08h:31m is saved in the files for 08h:30m and 08h:31m.

The main motivation for the DMS for the MUSER is to manage the metadata of each frame and provide the services of data query and retrieval from massive amounts of historical observational data with the query fields listed in Table~\ref{tab:query_condition}. Meanwhile, two key requirements are significant in designing the DMS.

1. High performance. As mentioned in the previous section, the MUSER will generate 23.04 million frames in an observational day lasting 10 hours and 8,409.6 million frames in an observational year of 3,650 hours; that is to say, the MUSER will generate approximately 1,281 TB observation data per year. If we use RDBMS to store the metadata of each frame, almost all mainstream database systems, i.e., MySQL, PostgreSQL, and Oracle, would struggle to support the efficient querying of a record or a range of records while dealing with so many records. These database systems commonly need to divide the data into several tables or sub-databases to improve performance. However, this throws up more difficulties in terms of programming. 

2. Small size. It is necessary to reduce the storage volume of the database as much as possible. Although all DMSs maintain the metadata of the observational data instead of the original raw data, the massive numbers of records occupy a large amount of storage space.

\begin{table*}[htbp]
    \centering
    \caption{Table of query fields}
    \begin{tabular}{c|c|c}
    \hline
        No & Field Name & Description \\
    \hline
        1 & start time &  The start date and time for query\\
    
        2 & end time  & The end date and time for query\\
    
        3 & band number &  Band number (e.g., 0: 400-800 MHz, 1:800-1200 MHz) \\
    
        4 & polarization & Polarization ( 0:Circular R, 1: Circular L) \\
        \hline
    \end{tabular}
    \label{tab:query_condition}
\end{table*}

\subsection{Negative Database For MUSER}

We have referred to the concept of negative representation of information based on the complement set \citep{esponda2004enhancing, halmos1960naive}, and propose a novel database technique, the negative database (ND).

The principle of the ND is simple. Given a universal set $F$, a dataset $A$, and its complement set $B$, if we can use $B$ to calculate $A$, the database of $B$ is a negative database. 
The creation of the ND assumes the following: 1) The universal dataset is known, and can be accurately defined. 2) All records can be derived or calculated based on an initial condition. 3) Dataset $A$ can be derived from its complement set $B$. 
Performance is a critical issue in the ND. 

\subsection{Design and Implementation} 
We choose the key-value database to implement the negative database, and construct a database management system for the MUSER. Given that Redis is a memory-based high performance key-value database, we use it as the underlying database system to implement the negative database on MUSER for massive data management. 

\subsubsection{Key-Value Design}

To maintain all required information,  two forms of key-value pairs are used. For presentation purposes, we use ``Key1-Value1" to express the first form and ``Key2-Value2" for the second. 

1. Key1-Value1

``Key1-Value1" is used to record the relation between the file and the observation date/time. The type of ``key1" is Redis string of length 12. The value of key1 is the observation date/time in the format  ``YYYYMMDDhhmm" (YYYY: Year, MM: Month, DD: day, hh: hour, mm: minute). The type of ``value1" is Redis set. 

For a specified key1, the element of value1 should store three parts consisting of the filename of the observation data file, the observation date and time ($T_1$) of the first frame, and the observation date and time ($T_{19200}$) extracted from the last frame. 

As mentioned in the previous section, data within one minute can be stored in two files. Therefore, to quickly locate the data in a minute with key1=``201702141011", two key-value pairs with the same key1 are added to the database.

2. Key2-Value2

``Key2-Value2" records detailed information concerning continuous frames and lost frames. The type of ``key2" is Redis string of length 20. The value of key2 is the observation date and time of the first frame in the format ``YYYYMMDDhhmmssffffff" (YYYY: Year, MM: Month, DD: day, hh: hour, mm: minute, ss: second, ffffff: microsecond). The type of ``value2" is Redis sorted set.

For a specified key2, the element of value2 should store five parts consisting of the start offset (S), the band number (B) extracted from the frame whose sequence number is the same as the start offset, the polarization (P) extracted from the frame whose sequence number is the same as the start offset, the end offset (E) and the cumulative amount (CA) of lost frames.

\begin{figure*}[htbp]
   \centering
   \includegraphics[width=0.8\linewidth]{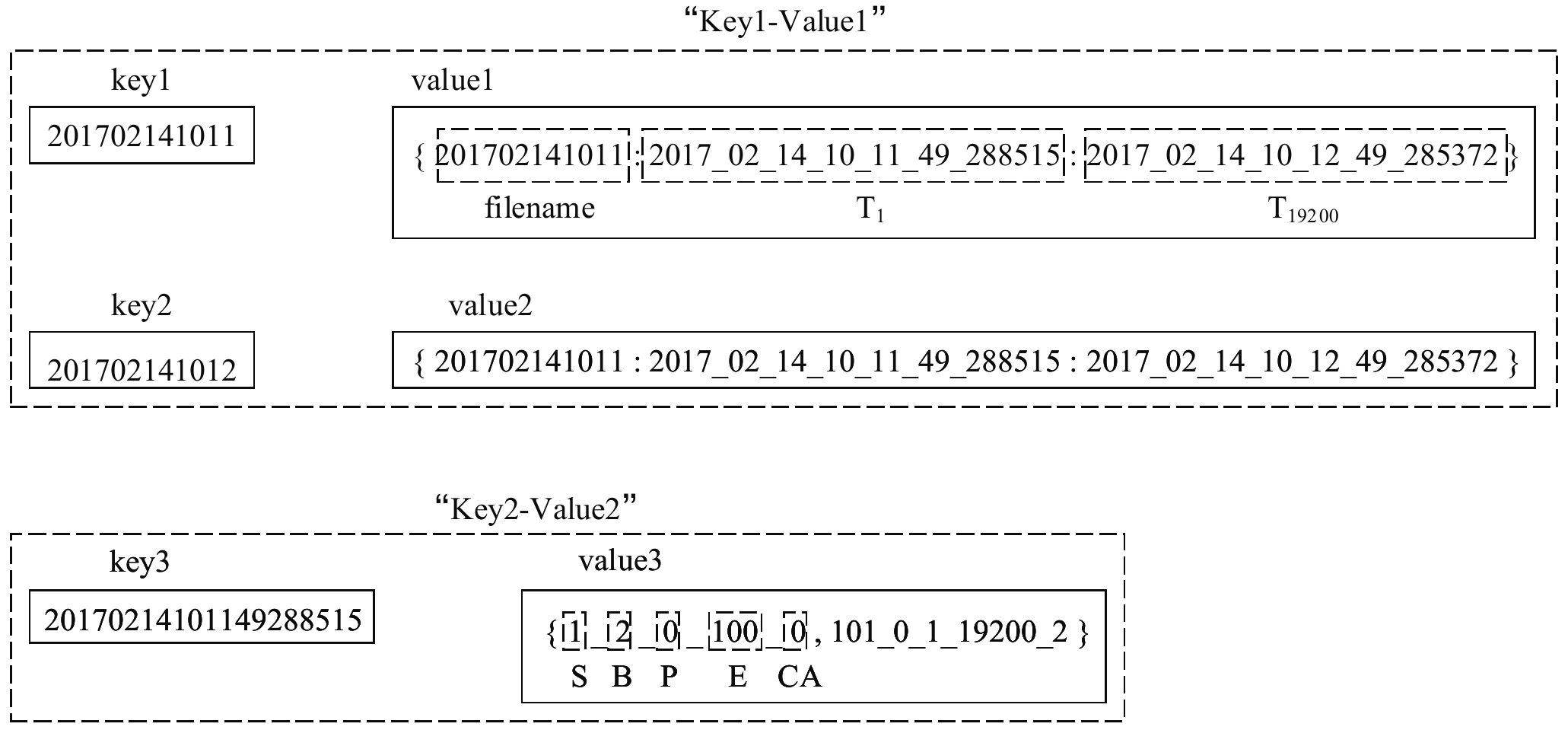}
   \caption{Two forms of key-value pairs generated by the file ``201702141011." }\label{FigureTwoForms}
   \end{figure*}

For example, the observation data file named ``201702141011," the observation date and time of the first and the last frames of which are ``20170214101149288515" and ``20170214101249285372," respectively, has one lost frame position between the frame whose sequence number is 100 and the frame with sequence number is 101, where there are two lost frames. The band number and polarization of the first and the 101th frames in ``201702141011" are ``2$\&$0" and ``0$\&$1," respectively. Figure~\ref{FigureTwoForms} shows the two forms of Key-Value pairs generated by the observation data file ``201702141011."

\subsubsection{Database Initialization}

\begin{figure*}[htbp]
   \centering
   \includegraphics[width=0.8\linewidth]{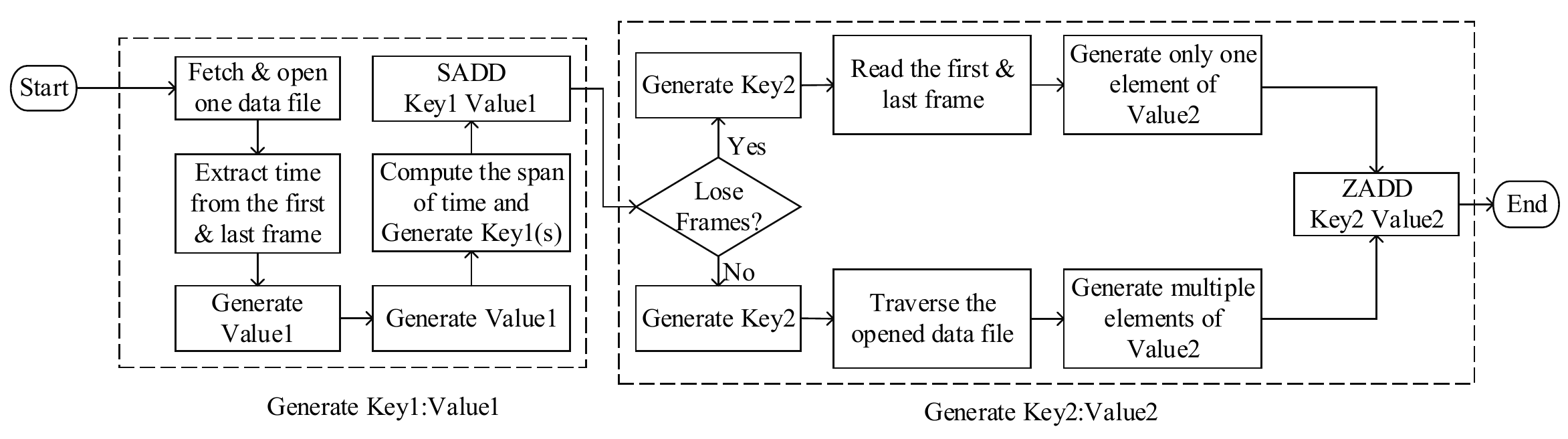}
   \caption{Flowchart of data initialization. }\label{FigureDataInit}
   \end{figure*}

Inserting all data is the first step of database initialization. A Python program was designed to scan the file(s), extract the information needed, generate the key-value pair, and insert it into the key-value database. Figure~\ref{FigureDataInit} shows a flowchart of data initialization for the creation of the negative database. 

In practice, the program is invoked following daily observation. It fetches all file names in the specified directory, and analyzes and extracts information from each file one by one. 

\subsubsection{Data Retrieval}

Data retrieval performance is one of the most important issues in database operation. Compared with RDBMS, data retrieval in the ND is quite complicated because many data records are deduced from existing key-value pairs. With such conditions such as observation date and time, data retrieval should output the corresponding frame(s) as quickly as possible.

   \begin{figure*}[htbp]
   \centering
   \includegraphics[width=0.8\linewidth]{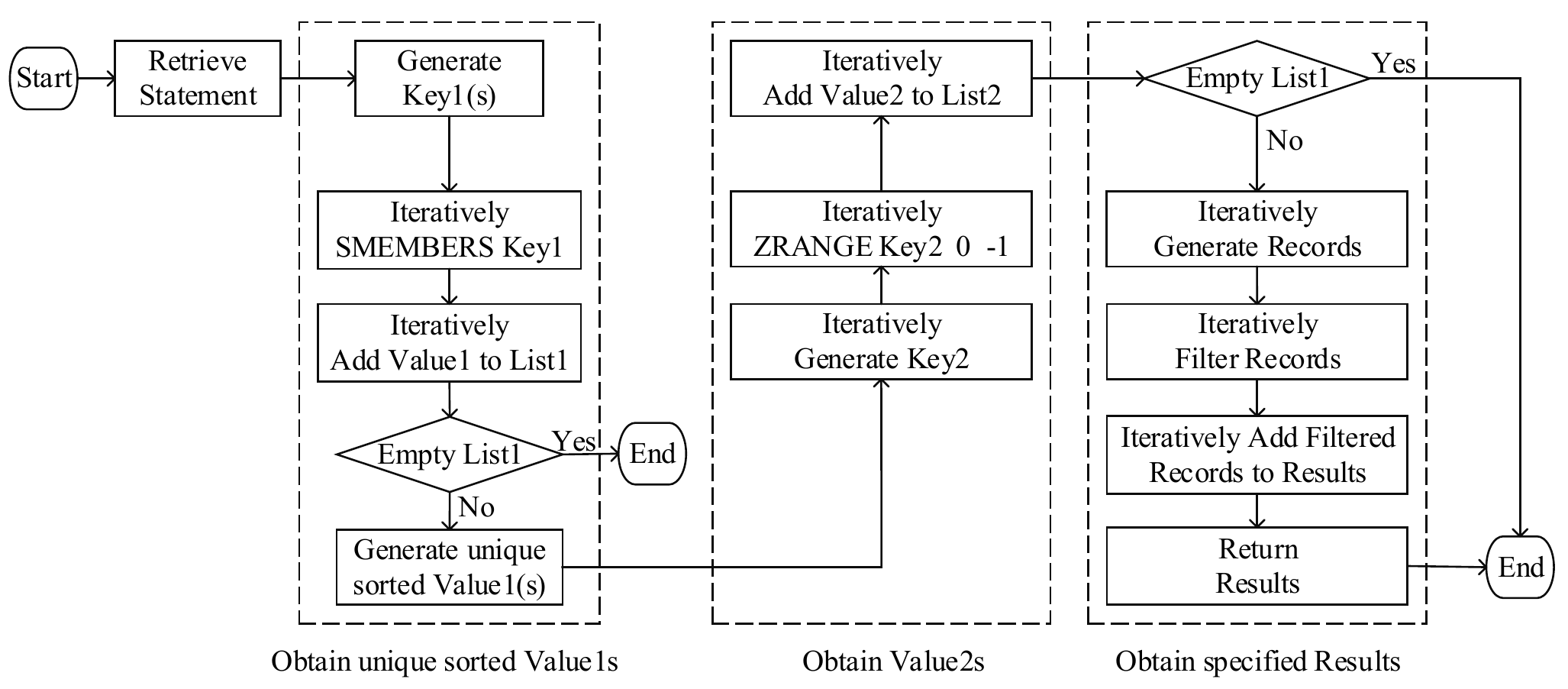}
   \caption{Flowchart of data retrieval. }\label{FigureDataRetr}
   \end{figure*}

Figure~\ref{FigureDataRetr} shows a flowchart of data retrieval of the negative database system for the MUSER. The full procedure can be described in three steps as follows: 

1. Obtaining unique sorted Value1s. The first step involves generating possible Key1(s) according to the time span computed using the specified retrieval time condition; in the second step, all possible repetitive unsorted Value1s are retrieved from the database by iteratively using the previously generated Key1(s); in the final step, we obtain a unique sorted Value1(s) in temporal order by using set theory and a sort algorithm.

2. Obtaining Value2(s). The first step involves generating Key2s by iteratively splitting Value1 into three parts and converting the second part into Key2; the second step consists of obtaining all Value2s from the database by iteratively using the generated Key2.

3. Obtaining specified results. The first step consists of iteratively generating possible intermediate records in one observational data file according to the cyclic mode and the elements of Value2; the second step is to iteratively filter the partially specified results from the generated intermediate records according to the specified retrieval conditions, and iteratively add the filtered records to the results; the third step involves returning the specified results.

\section{Performance Test}

Two kinds of tests were designed to test the performance of the ND based on MySQL using the MyISAM engine with the default key buffer (key cache) size of 8 MB,  default key cache block size of 1KB, 300 as the default value of the key cache age threshold, 100 as the default value of the key cache division limit, and Redis using the default configuration. The first test compared the performance of the proposed system with that of the RDBMS technique on real observational data. The second tested performance variations in both systems under different frame loss ratios.

The experimental platform was a PC server with two-way Intel Xeon E5-2650 v2 CPUs, 2.6 GHz, 16 cores, 32 GB memory, and a 1 TB hard disk. It ran on the CentOS 6.8 operating system with the MySQL database version 5.7.18. The key-value database of Redis used was latest stable version, 3.2.6, and Python version 2.7.10 was used as well.

\subsection{Performance Comparison}

In order to test the performance of the ND based on MySQL and Redis, we created a relational database based on MySQL with the configuration parameters referring to the previous section 4.

As shown in Table~\ref{tableschema}, we created five fields, such as observational date and time, file name, polarization, band, and the offset of the frame in bytes, for the database of the DMS. We extracted the metadata from the frames using 406 real observational data files generated by the MUSER-I on Nov 1, 2015 and these 406 observational data files represented approximately 6.8 hours of observation. A total of 7,795,200 records were appended to the database as test data. A total of 20 observational data files had randomly lost some frames, with at most 10 such instances in each data file, and at most four frames were lost at each position.

\begin{table*}[htbp]
    \centering
    \caption{Description of table schema}
    \begin{tabular}{c|c|c|c|c}
    \hline
        Field & Type & Null & Key & Extra \\
    \hline
        observationalTime & bigint(64) & NO  & PRI & \\
    
        filename & varchar(100) & YES &  & \\
    
        polarization & bit(1) & YES &  &\\
    
        band & tinyint(8) & YES &   &\\
    
        offset & int(11) & YES &   &\\
        \hline
    \end{tabular}
    \label{tableschema}
\end{table*}

Two tests were conducted. The first compared the data initialization performance of the ND with that of RDBMS, and the second compared the data retrieval performance of the two.

\subsubsection{Database Initialization Performance}\label{DB-Init-Performance}

\begin{figure*}[htbp]
   \centering
   \includegraphics[width=0.8\linewidth]{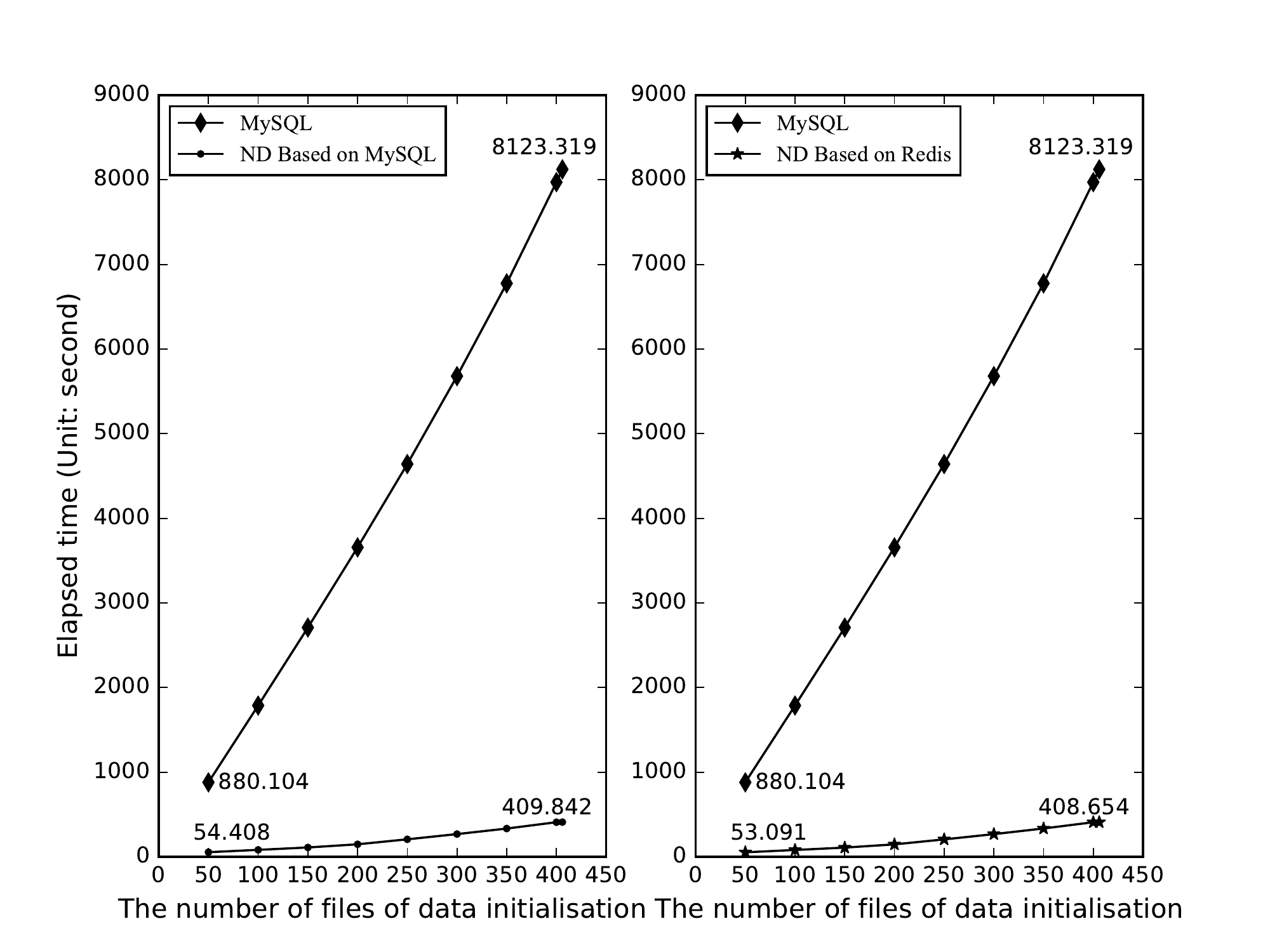}
   \caption{Comparison of data initialization performance between ND and RDBMS. }\label{FigureInitPerformance}
   \end{figure*}

To guarantee accuracy in our comparison, we repeated the test comparing database initialization performance six times; the average data initialization times were calculated, and are plotted in Figure~\ref{FigureInitPerformance}. As is evident from the figure, for the 406 observational data files, the ND was nearly 20 times faster than RDBMS. It recorded an average time one second for each observational data file, whereas that for RDBMS was close to 20 s. The ND based on Redis was slightly faster than the ND based on MySQL.

The MySQL database contained 7,795,200 rows and took up 843.3 MB of storage consisting of 107.4 MB of indices and 739.9 MB of data. Whether based on Redis or MySQL, the ND database only contained 812 records and occupied approximately 153 KB. The storage volume occupied by RDBMS was approximately 5,644 times greater than that occupied by the ND. The records generated by RDBMS were 9,600 times as many as those by the ND.

\subsubsection{Data Retrieval Performance}

We tested the performance for retrieving 1 (3 ms), 8 (25ms, 1 full frame), 80 (250 ms), 160 (500 ms), 320 (1 s), and 640 (2 s) consecutive frame(s) from the database. These numbers were chosen to meet the requirements of visibility data integral calculation. 
Scientists often overlay (integral) the visibility data from several frames to increase the signal-to-noise ratio (SNR). The number of integral calculation times are 0, 0, 10, 20, 40 and 80. 

\begin{figure*}[htbp]
   \centering
   \includegraphics[width=0.8\linewidth]{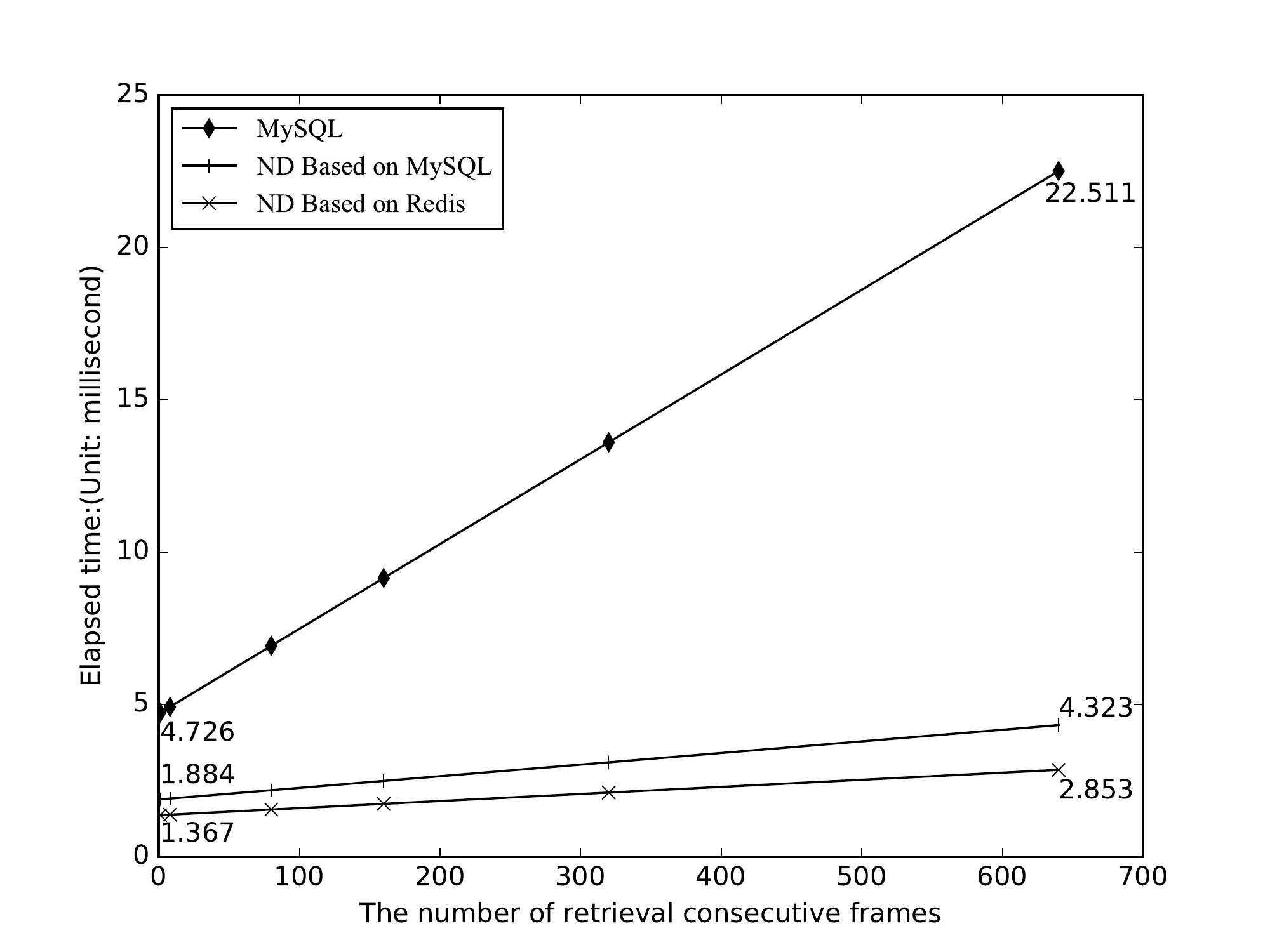}
   \caption{Comparison of data retrieval performance between ND and RDBMS. }\label{FigureRetrPerf}
   \end{figure*}

To guarantee the accuracy of the tests, we repeated them 1,000,000 times and recorded the response times for data retrieval. The final result consisted of the average values of these test records, which are shown in Figure~\ref{FigureRetrPerf}. 

The data retrieval performance of the ND was faster than that of RDBMS.
Moreover, with increasing number of frames to be retrieved, the performance of RDBMS significantly suffered, whereas that of the ND was only slightly affected. In other words, with increasing number of frames, data retrieval became much more difficult for RDBMS. At the same time, the data retrieval performance of the ND based on Redis was slightly faster than that based on MySQL.

\subsection{Performance Variation With Varying Frame Loss Ratios}

The retrieval performance of the ND is significantly dependent on the frame loss ratio of the observation data. In theory, a higher frame loss ratio leads to poorer performance.
We designed five groups of experimental data to test performance variation with different loss ratios. The frame loss ratios used were 0.25\%, 0.50\%, 1\%, 3\%, and 5\%. We assumed two lost frames at every instance of frame loss.
   
\begin{figure*}[htbp]
   \centering
   \includegraphics[width=0.8\linewidth]{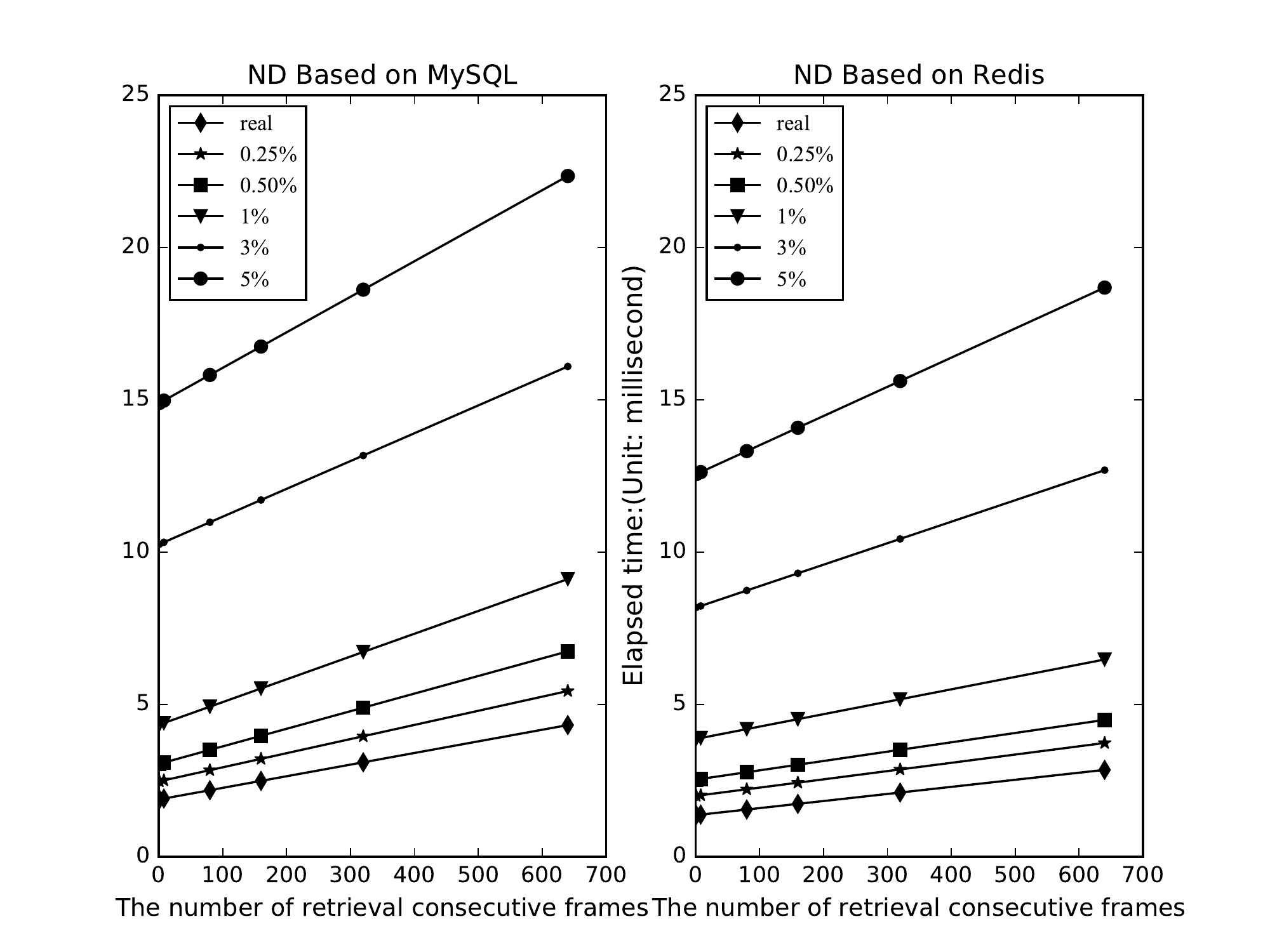}
   \caption{Data retrieval performance corresponding to different frame loss ratios. }\label{FigureDifferenLoss}
   \end{figure*}

As in the previous experiment, we repeated this one 1,000,000 times and recorded the data retrieval response times. The final average results are shown in Figure~\ref{FigureDifferenLoss}.

For frame loss ratios of 0.25\% and 0.50\%, the data retrieval response time increased one to two times compared with that for real observational data. However, for the other three groups of simulated data at a frame loss ratio of 1\%, 3\%, and 5\%, the data retrieval response time increased four to 10 times more than that for real observational data. Meanwhile, the ND based on Redis was slightly faster than the ND based on MySQL.

\section{Discussion}
\subsection{Performance}
The ND is a novel technique for massive data management. According to the experiment results, the carefully designed codes of the ND exhibited acceptable retrieval/query performance. The volume of required storage in the ND database was also significantly reduced. In this regard, the ND can be thought of as an effective technique for managing massive amounts of observational data generated by next-generation telescopes.

1. Database Initialization performance

The data size of the ND was nearly four orders of magnitude smaller than that of the RDBMS, but the initialization of the former was approximately 20 times faster than that of the latter (see Figure~\ref{FigureInitPerformance}). The high performance at data initialization of the ND was determined by the number of lost frame(s) in the file. The ND spent on average 20.14 seconds on an observation data file with lost frame(s) because it needed to parse the full file in such cases. If all files had lost frame(s), the data initialization performance of the ND and the RDBMS would have been more or less identical.

To improve the initialization performance of the ND, we should determine the existence of the lost frame(s) before inserting the records. It is easy to only need to read out the times of the first and last frames in a file. There must be lost frame(s) if the time difference is not one minute.

2. Retrieval performance

According to the experimental results, the retrieval performance of the ND is comparable to currently available RDBMSs. The retrieval performance of the ND consists of two parts: record retrieval performance, and the performance in deriving absent records. Although the complex procedure of deriving the final records is time-consuming, the record retrieval performance of the ND was eight times faster than that of the RDBMS (see Figure~\ref{FigureRetrPerf}).

3. Frame loss ratio

The frame lose ratio of the MUSER was less than 0.25\%. However, we noticed that the frame loss ratio was an important factor affecting the performance of the ND.  Figure~\ref{FigureDifferenLoss}) shows that the decline in retrieval performance was essentially proportional to the frame loss ratio. The reason is quite simple.  More records needed to be inserted into the complement set, because of which the derivation of records consumed considerably more time. When the frame loss ratio reached 5\%, the retrieval performance of the ND was inferior to that of the RDBMS. 

4. Redis and MySQL

The ND can run on many database systems.
In this study, we have tried Redis and MySQL respectively. According to the experimental results, both databases are suitable for creating a high performance negative database. Moreover, the retrieval performance of the ND based on Redis was slightly faster than that of the ND based on the MySQL. 

\subsection{Limitations}
There are some limitations to the ND: 

1. The current ND is simply a prototype that provides the fundamental data retrieval function. It can support the real applications of the MUSER data processing system. However, from the point of view of the database, it cannot support record updates and deletion, which needs to be implemented in the future work.

2. The principle underlying the ND determines that the data must be arranged consecutively at fixed time intervals. Otherwise, there is no way to derive the corresponding data based on a few discrete records. Therefore, the application and popularization of the ND is significantly limited. Fortunately, however, scientific data management is suited to the ND technique, since scientific data, e.g., astronomical radio data, are generally time-series data. These data have stable and fixed sampling intervals. Thus, the ND can be used for massive data management in next-generation telescope systems.

\section{Conclusion}

In this paper, we proposed a novel database technique, the negative database, to effectively manage massive scientific time-series data. Experimental results showed that the ND yielded acceptable performance in database initialization and data retrieval, and significantly reduced the volume of storage in the database. 

Although the ND has limitations, it can be effectively used in astronomy. This study can serve as a valuable reference for research on data management for next-generation telescopes.

\begin{acknowledgements}
This work is supported by the National Key Research and Development Program of China (2016YFE0100300) and the National Natural Science Foundation of China (No. U1231205, U1531132, U1631129,11403009, 11463003). The authors also gratefully acknowledge the helpful comments and suggestions of the reviewers.
\end{acknowledgements}



\bibliography{bibtex}
\bibliographystyle{aasjournal}



\end{document}